\documentclass[a4paper]{jpconf}
\usepackage{graphicx}
\usepackage{amsmath}
\usepackage{amssymb}
\usepackage{subfigure}
\usepackage[colorlinks]{hyperref}
\hypersetup{citecolor=blue, linkcolor=blue, urlcolor=blue}
\begin{document}

\title{Numerical Brill-Lindquist initial data with a Schwarzschildean end at spatial infinity}

\author{Georgios Doulis$^1$ and Oliver Rinne$^{1,2}$}

\address{$^1$ Max Planck Institute for Gravitational Physics (Albert Einstein Institute), Am M\"uhlenberg 1, 14476 Golm, Germany} 
\address{$^2$ Department of Mathematics and Computer Science, Freie Universit\"at Berlin,
Arnimallee 26, 14195 Berlin, Germany}

\ead{georgios.doulis@aei.mpg.de, oliver.rinne@aei.mpg.de}

\begin{abstract}
We construct numerically time-symmetric initial data that are Schwarzschildean at 
spatial infinity and Brill-Lindquist in the interior. The transition between these 
two data sets takes place along a finite gluing region equipped with an axisymmetric 
Brill wave metric. The construction is based on an application of Corvino's gluing 
method using Brill waves due to Giulini and Holzegel. Here, we use a gluing function 
that includes a simple angular dependence. We also investigate the dependence of 
the ADM mass of our construction on the details of the gluing procedure. 
\end{abstract}

\section{Introduction}
\label{sec:intro}

The long lasting question of whether it is possible to construct solutions to 
the vacuum constraint equations that are static in a neighbourhood of space-like 
infinity and non-static in the interior was answered to the affirmative by Corvino 
\cite{Corvino2000}. Therein, Corvino proved that an external Schwarzschild region 
can be glued smoothly along a finite intermediate annulus to generic asymptotically 
flat time-symmetric data in the interior. This result paved the way to its generalisations 
to initial data with stationary Kerr \cite{Corvino2006} or Kerr-de Sitter \cite{Cortier2013} 
ends and static Kottler-de Sitter \cite{Chrusciel2008} or Kottler-anti-de Sitter 
\cite{Chrusciel2009} ends. Although the above analytical work provides results 
about the existence of the data sets considered there, it is not very explicit 
in the sense that it is not clear how to construct these data explicitly through 
analytical or numerical methods. The first step, to our knowledge, towards this 
direction was taken in \cite{Giu&Hol2005}, where the authors proposed an explicit 
formulation of the gluing procedure for vacuum axisymmetric initial data. Recently, 
we presented the first attempt to implement this scheme numerically \cite{Dou&Rinn2014}. 
In the present short contribution, we focus on a setting where the gluing function 
acquires a simple angular dependence.

The motivation to construct such kind of solutions to the vacuum constraint equations 
comes from the fact that 
initial data of static or stationary character close to spatial infinity seem 
to have a smooth development to null infinity \cite{Chrusciel2002}---a result 
of great importance if one is interested in extracting unambiguously information 
about the gravitational radiation emitted by isolated self-gravitating systems. 
In order to evolve the aforementioned initial data, one could use a combination 
of Cauchy and hyperboloidal evolution  in the following sense. For example, in 
the static case the initial data are Schwarzschild outside the gluing annulus, 
thus one can place an artificial boundary there and evolve the data for a short 
time using a standard Cauchy evolution with exact Schwarzschild boundary conditions 
being imposed on the artificial boundary. Then, these evolved data can be used 
as initial conditions for a hyperboloidal evolution \cite{Rinne2010}. Alternatively, 
the initial data could be evolved using a Cauchy evolution in the framework of 
the conformal representation of Einstein's equations \cite{Friedrich1998}, where 
space-like infinity $i^0$ has been blown up to a cylinder $I = [-1, 1] \times \mathbb{S}^2$ 
of finite length along the time direction. This approach has been already proven 
to be successfully numerically implementable, e.g. \cite{Beyer2012,Doulis2013}.

\section{Mathematical formulation}
\label{sec:formulation}

We intend to construct initial data that in the interior region $0 < r < r_\mathrm{int}$ 
consist of two identical non-rotating black holes of mass $m$ lying symmetrically 
to the origin on the z-axis, $\vec{c} = (0, 0, \frac{d}{2})$, at a moment of time 
symmetry:\footnote{Here only black holes with non-intersecting horizons will be 
considered and all cases that a third outer horizon enclosing both black holes 
appears will be excluded.}
\begin{equation}
 \label{B-L_metric}
 g_{\textrm{\tiny B-L}} = \left(1 + \frac{m}{2|\vec{r} - \vec{c}|} + \frac{m}{2|\vec{r} + \vec{c}|} \right)^4 
 \delta,
\end{equation}
where $\delta = dr^2 + r^2 (d\theta^2 + \sin^2\theta\, d\phi^2)$ denotes the three-dimensional 
Minkowski line element in spherical polar coordinates. In the intermediate gluing 
region $r_\mathrm{int} < r < r_\mathrm{ext}$ the data are of the general Brill wave 
form
\begin{equation}
 \label{Brill_metric}
 g_{\textrm{\tiny Brill}} =  \psi^4 \left(e^{2\,q(r,\theta)} (dr^2 + r^2 d\theta^2) + r^2 \sin^2\theta\, d\phi^2\right),
\end{equation}
and in the exterior region $r > r_\mathrm{ext}$ they are Schwarzschild,
\begin{equation}
 \label{Schw_metric}
 g_{\textrm{\tiny Schw}} = \left(1 + \frac{M}{2|\vec{r}|} \right)^4 \delta,
\end{equation}
with ADM mass $M$. Following \cite{Giu&Hol2005}, one can glue together the above 
data sets by defining the conformal factor
\begin{equation}
\label{conf_factor}
 \psi = \left(1 + \frac{m}{2|\vec{r} - \vec{c}|} + \frac{m}{2|\vec{r} + \vec{c}|} \right) \beta(r,\theta) 
 +(1 - \beta(r,\theta)) \left(1 + \frac{M}{2|\vec{r}|} \right).
\end{equation}
Whereas in \cite{Dou&Rinn2014} we also considered a $\theta$-independent gluing 
function $\beta$, we focus here on a gluing function with a simple $\theta$-dependence, 
namely we will assume that 
\begin{equation}
  \label{beta_function}
  \beta(r, \theta) = \frac{1}{2} \left(1 + \tanh \left(\frac{1}{r - r_{\mathrm{int}}} 
   + \frac{1}{r - r_{\mathrm{ext}}}\right)\right) + \mathrm{sech} \left(\frac{1}{r - r_{\mathrm{int}}}
   + \frac{1}{r - r_{\mathrm{ext}}}\right) \sin^2\! \theta.
\end{equation}
Notice that $\beta(r \leq r_\mathrm{int}, \theta) = 1$ and $\beta(r \geq r_\mathrm{ext}, \theta) = 0$.

The Brill wave function $q(r,\theta)$ will be our unknown here and will be specified 
by Einstein's equations, which in our setting reduce to the vanishing of the Ricci 
scalar of the Brill wave metric \eqref{Brill_metric}, i.e. $R({g}_{\textrm{\tiny Brill}}) = 0$. 
After a little bit of algebra the latter reads
\begin{equation}
\label{Poisson_eq}
 \frac{\partial^2 q}{\partial r^2} + \frac{1}{r^2} \frac{\partial^2 q}{\partial \theta^2} + 
 \frac{1}{r}  \frac{\partial q}{\partial r} = 
 -\frac{4}{\psi} \left( \frac{\partial^2 \psi}{\partial r^2} + \frac{1}{r^2} \frac{\partial^2 \psi}{\partial \theta^2} +
 \frac{2}{r} \frac{\partial \psi}{\partial r} + \frac{\cot\theta}{r^2} \frac{\partial \psi}{\partial \theta} \right).
\end{equation}
Notice that the right-hand side of the elliptic equation \eqref{Poisson_eq} is 
known a priori as the conformal factor \eqref{conf_factor} and the gluing function 
\eqref{beta_function} are given. The inhomogeneous Poisson equation \eqref{Poisson_eq} 
will be supplemented by the boundary conditions
\begin{subequations}
 \label{bound_cond}
  \begin{align}
   &q = 0 \quad \mathrm{and} \quad \frac{\partial q}{\partial \theta} = 0 \qquad \,\,\, \mathrm{at} \quad 
   \theta = 0, \pi,  \label{bound_cond_theta}\\
   &q = 0 \quad \mathrm{and} \quad \frac{\partial^n \! q}{\partial r^n} = 0 \qquad \mathrm{at} \quad
   r = r_{\mathrm{int}}, r_{\mathrm{ext}}  \label{bound_cond_r},
  \end{align}
\end{subequations}
for all $n \in \mathbb{N}$. Conditions \eqref{bound_cond_theta} guarantee the 
absence of any conical singularities on the z-axis, while \eqref{bound_cond_r} 
ensure that the transition between the different data sets at the boundaries 
$r_\mathrm{int}, r_\mathrm{ext}$ of the gluing annulus is smooth. 

It is worth mentioning that by inserting into \eqref{Poisson_eq} the definition 
of the conformal factor \eqref{conf_factor}, the former takes the form
\begin{equation*}
  \frac{\partial^2 q}{\partial r^2} + \frac{1}{r^2} \frac{\partial^2 q}{\partial \theta^2} + 
 \frac{1}{r} \frac{\partial q}{\partial r} = 
 \frac{1}{M + 2\,r + f_0\, \beta} \left(f_1\, \frac{\partial^2 \beta}{\partial r^2} + f_2\, \frac{\partial^2 \beta}{\partial \theta^2} +
 f_3\, \frac{\partial \beta}{\partial r} + f_4\, \frac{\partial \beta}{\partial \theta} \right),
\end{equation*}
where $f_k$ are functions of $r$ and $\theta$. Notice that the constancy of 
the gluing function \eqref{beta_function} outside the gluing annulus forces 
the right-hand side of the above equation to vanish. In addition, the boundary 
conditions \eqref{bound_cond_r} guarantee that $q$ is zero outside the gluing 
region. Thus, the Poisson equation \eqref{Poisson_eq} reduces to the trivial 
identity ($0 = 0$) in the exterior of the gluing annulus. 

In summary, our goal in the following section will be to solve numerically 
the elliptic second-order PDE \eqref{Poisson_eq} for $q(r, \theta)$ subject 
to the boundary conditions \eqref{bound_cond}.

\section{Numerical implementation}
\label{sec:implementation}

\subsection{Numerical scheme}
\label{sec:scheme}

We use pseudo-spectral methods to solve numerically the system \eqref{Poisson_eq}, \eqref{bound_cond}. 
First, we perform the transformation 
$x \mapsto r(x) := \frac{1}{2}(r_{\mathrm{ext}} - r_{\mathrm{int}})\, x + \frac{1}{2}(r_\mathrm{ext} + r_\mathrm{int})$ 
on the radial coordinate in order to map the physical domain $(r, \theta) \in [r_\mathrm{int}, r_\mathrm{ext}] \times [0, \pi]$ 
to the computational domain $(x, \theta) \in [-1, 1] \times [0, \pi]$, which will enable us 
to expand the radial dependence of $q$ in Chebyshev polynomials $T_k$ and the $\theta$-dependence 
in Fourier-sine series. We discretize the computational domain by introducing the collocation 
points along the angular and the new radial direction
\begin{equation*}
 \theta_i = \frac{i\, \pi}{L} \qquad \mathrm{and} \qquad x_j =  -\cos\left(\frac{j\, \pi}{K}\right) 
 \qquad \mathrm{with} \qquad i = 0, \ldots, L \quad \mathrm{and} \quad j = 0, \ldots, K,
\end{equation*}
where $L$ and $K$ are the numbers of collocation points.

To satisfy the boundary conditions \eqref{bound_cond} we make the following ansatz for the 
Brill wave function,
\begin{equation}
\label{ansatz}
 q(x, \theta) = B(x) \sin\theta\, \sum^K_{k=0}{\sum^{L-1}_{l=1}{a_{kl}\, T_k(x)\, \sin(l\, \theta)}},
\end{equation}
where $K, L$ as above, the constants $a_{kl}$ are the expansion coefficients in our series 
expansion, and $B(x)$ is a ``bump'' function of the form
\begin{equation}
\label{bump_function}
 B(x) = \mathrm{sech} \left(\frac{b_1}{x - 1} + \frac{b_2}{x + 1} \right) \quad \textrm{with} \quad b_1, b_2 \quad \textrm{constants}.
\end{equation}
While the presence of $B(x)$ in \eqref{ansatz} guarantees that the set of boundary 
conditions \eqref{bound_cond_r} is satisfied, the expansion in Fourier-sine series 
and the presence of $\sin\theta$ ensures that our numerical solutions will respect 
the conditions \eqref{bound_cond_theta} on the z-axis.

\subsection{Numerical solutions}
\label{sec:solutions}

In order to produce numerical solutions for the system \eqref{Poisson_eq}, \eqref{bound_cond} 
using the method described in Sec.~\ref{sec:scheme}, one has first to express \eqref{Poisson_eq} 
in terms of the new radial coordinate $x$ and then substitute the ansatz \eqref{ansatz} 
and the definitions \eqref{conf_factor}, \eqref{beta_function} into the left and right-hand 
side, respectively, of the resulting expression. In addition, one has to specify the 
free parameters entering the definitions of the conformal factor \eqref{conf_factor} 
and the gluing function \eqref{beta_function}, namely the mass $m$ of the individual 
Brill-Lindquist black holes, their mutual distance $d$, the mass $M$ of the Schwarzschild 
region, and the locations of the boundaries $r_\mathrm{int}, r_\mathrm{ext}$ of the 
gluing annulus. There are three conditions that constrain the choice of the free parameters: 
$i)$ the mass-to-distance ratio $m/d \leq 0.64$ that prevents the occurence of a third 
outer horizon enclosing the Brill-Lindquist data, see \cite{Bril&Linq1963}, $ii)$ the 
inequality $r_\mathrm{int} > d/2 + m\,d/(2\,d + m)$ that guarantees that the gluing 
annulus is placed outside of any horizons of the Brill-Lindquist data \cite{Dou&Rinn2014}, 
and $iii)$ the integrability condition presented in Sec.~\ref{sec:adm_mass} that constrains 
the relation between the masses $m$ and $M$.

\begin{figure}[htb]
 \centering
  \subfigure[]{
   \includegraphics[scale = 0.37]{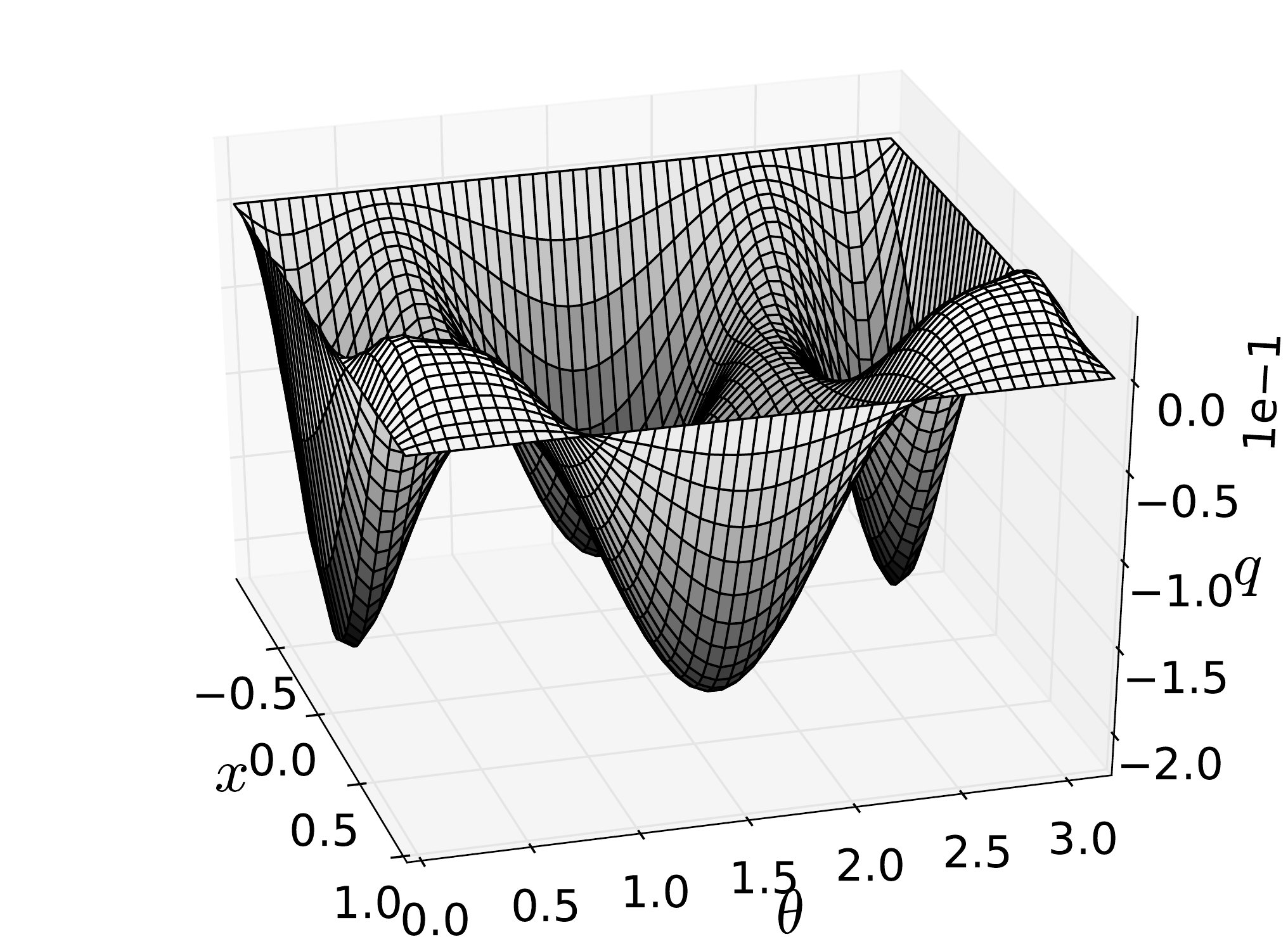} 
   \label{fig:num_sol_a}
  }
  \subfigure[]{
   \includegraphics[scale = 0.37]{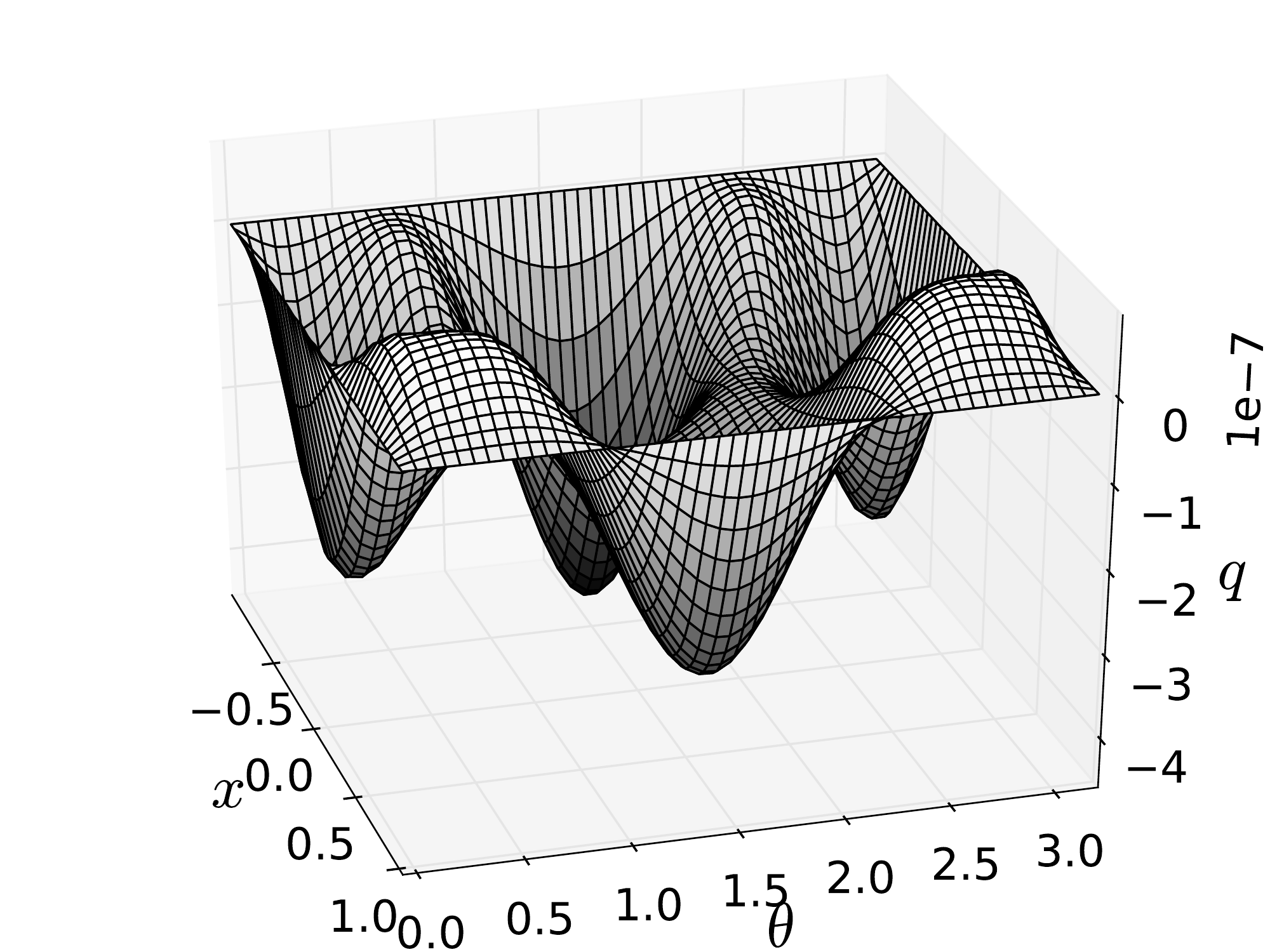} 
   \label{fig:num_sol_b}
  }
 \caption{Numerical solutions. The numerical values of the function $q$ for the choice 
 $m = 2$, $d = 10$, $r_{\mathrm{ext}} = 2\, r_{\mathrm{int}}$, and $M$ according to 
 Tab.~\ref{tab:ADM_increase} with the gluing annulus placed at (a) $r_{\mathrm{int}} = 6$ 
 and (b) $r_{\mathrm{int}} = 500$. The values of $q$ decrease with the distance of the 
 gluing annulus from the origin. The resolution used to produce the plots is $K=L=35$ 
 collocation points.
 }
 \label{fig:num_solution}
\end{figure}

In Fig.~\ref{fig:num_solution} numerical solutions of the system \eqref{Poisson_eq}, 
\eqref{bound_cond} are presented for a choice that respects the above three conditions, 
i.e. $m = 2$, $d = 10$, $r_\mathrm{ext} = 2\, r_\mathrm{int}$, and $M$ takes values 
in accordance with Tab.~\ref{tab:ADM_increase}. We provide solutions for two different 
locations of the gluing annulus. In Fig.~\ref{fig:num_sol_a} the gluing region has 
been placed very close to the black holes ($r_\mathrm{int} = 6$), in Fig.~\ref{fig:num_sol_b}
far away ($r_\mathrm{int} = 500$). The numerically computed values of $q$ drop with 
the distance of the gluing annulus from the origin. This behaviour was also observed 
in \cite{Dou&Rinn2014} for solutions corresponding to a $\theta$-independent gluing 
function. 

The accuracy and convergence properties of the solutions presented in Fig.~\ref{fig:num_solution} 
are similar to the ones observed in \cite{Dou&Rinn2014}. Namely, the expansion coefficients 
in \eqref{ansatz} decay to zero exponentially, and their actual values $a_{kl}$ for 
$k, l > 60$ are of the order of $10^{-14}$ (numerical roundoff).

\subsection{Behaviour of the ADM mass}
\label{sec:adm_mass}

Here, we will investigate any possible effects that the introduction of the gluing 
region has on the total ADM mass $M$ of our construction. One expects that the 
presence of the Brill wave in the gluing region would in general increase the 
overall ADM mass of the gluing construction. As was pointed out in \cite{Dou&Rinn2014}, 
the positive mass theorem for time-symmetric, axisymmetric, vacuum gravitational 
waves proved by Brill \cite{Brill1959} can be used in the form 
\begin{equation}
 \label{integr_cond}
  M = \int^\pi_0 \int^\infty_0 \left[ \left( \frac{1}{\psi}\frac{\partial \psi}{\partial r} \right)^2 + 
  \left( \frac{1}{r\, \psi}\frac{\partial \psi}{\partial \theta} \right)^2 \right] r^2 \sin\theta\, dr\, d\theta,
\end{equation}
not only to determine the relation between the masses $m$ and $M$ involved in our 
construction, but also to study the behaviour of the ADM mass $M$ on the details 
of the gluing construction. In the expression above $\psi$ is the conformal factor 
defined by \eqref{conf_factor}, thus the ADM mass $M$ enters also the right-hand 
side of \eqref{integr_cond}. In the light of this observation, the relation \eqref{integr_cond} 
can be viewed as an integrability condition for the ADM mass in the following sense. 
For a specific choice of $M$, we compute numerically the integral quantity on the 
right-hand side of \eqref{integr_cond} and denote the result with $M_I$. We will 
be interested here in the cases that $M_I = M$ holds, as they correspond to true 
physical solutions. The case that $M_I$ takes values $M_I < 2 m$ corresponds to 
a reduction of the ADM mass, the case $M_I > 2 m$ to an increase.

In Tab.~\ref{tab:ADM_increase} a detailed study of the restrictions that the condition 
\eqref{integr_cond} poses on the ADM mass is presented for the choice $m = 2$, $d = 10$, 
$r_\mathrm{ext} = 2\, r_\mathrm{int}$ of the free parameters that was used to generate 
the solutions of Fig.~\ref{fig:num_solution}. The closer we place the gluing annulus 
to the origin, the greater is the increase of the ADM mass. The increase is limited 
only by the requirement that the gluing region lies outside of the horizons of the 
Brill-Lindquist data, i.e. $r_\mathrm{int} > d/2 + m\,d/(2\,d + m)$. Notice that 
the actual increase of the ADM mass decreases extremely fast to zero with increasing 
gluing radius $r_\mathrm{int}$.
\begin{table}[htb]
 \centering
 \begin{tabular}{|c|c|c|c|c|}
  \hline
  $r_\mathrm{int}$    &         $M_I$        &     $M_I - 2 m$           &       $|\Delta M|$       &        $\delta M$        \\ \hline
       6              &       4.08778        &      0.08778              &  $1.4 \times 10^{-5}$    &   $1.5 \times 10^{-5}$   \\ \hline
       10             &       4.0119         &      0.0119               &  $3.9 \times 10^{-5}$    &   $4.5 \times 10^{-5}$   \\ \hline
       30             &       4.00017185     &      0.00017185           &  $3.4 \times 10^{-8}$    &   $7.9 \times 10^{-8}$   \\ \hline
       100            &       4.00000145     &      $1.45 \times 10^{-6}$&  $1.6 \times 10^{-8}$    &   $2.2 \times 10^{-8}$   \\ \hline
       500            &       4.0000000024   &      $2.4 \times 10^{-9}$ &  $3.3 \times 10^{-11}$   &   $4 \times 10^{-11}$    \\ \hline
 \end{tabular}
 \caption{Increase of the ADM mass. The increased ADM mass, the actual increase, 
 the violation of the integrability condition $\Delta M = M_I - M$, and the 
 numerical error $\delta M$ involved in these calculations are presented for 
 different locations of the gluing annulus. Notice that by just quintupling 
 the gluing radius (from 6 to 30), the increase decreases by two orders of 
 magnitude.}
 \label{tab:ADM_increase}
\end{table}

In \cite{Dou&Rinn2014} it has been shown that for the $\theta$-independent gluing 
function 
\begin{equation*}
\beta(r) = \frac{1}{2} \left(1 + \tanh \left(\frac{1}{r - r_{\mathrm{int}}} + \frac{1}{r - r_{\mathrm{ext}}}\right)\right), 
\end{equation*} 
it is possible to reduce the ADM mass in the rather extreme case that the black 
holes are placed very close to each other and the gluing is performed along a 
very wide gluing annulus that starts very close to the horizons of the black 
holes. Here, it was not possible to find a setup that could lead to a reduction 
of the ADM mass for the choice \eqref{beta_function}, even for such extreme 
configurations. It seems that even the simplest inclusion of $\theta$-dependence 
in the gluing function increases the contribution of the Brill wave to the ADM 
mass to a degree that makes it impossible to find an arrangement of our data 
that leads to reduction of the ADM mass.

\section{Conclusions}

In \cite{Dou&Rinn2014} we presented a numerical implementation of Corvino's 
gluing method \cite{Corvino2000} applied to vacuum axisymmetric spacetimes 
\cite{Giu&Hol2005}. Here, we focused on a gluing function with a simple $\theta$-dependence. 
It was shown numerically that for the choice \eqref{beta_function}, the vacuum 
time-symmetric constraint equations \eqref{Poisson_eq} admit smooth solutions 
that are of Schwarzschild type close to spatial infinity, see Fig.~\ref{fig:num_solution}. 
We showed that these solutions converge exponentially and depend on the location 
of the gluing annulus in the expected way: the Brill wave function $q$ decreases 
with the distance of the gluing annulus from the origin. The latter observation 
confirms the expectation that in the limiting case that the gluing annulus is 
placed at infinity $q$ must tend to zero. 

Our study of the dependence of the total ADM mass $M$ on the details of the 
gluing procedure carried out in Sec.~\ref{sec:adm_mass} strongly indicates 
that for the choice \eqref{beta_function} the presence of the gluing annulus 
tends always to increase the ADM mass. It seems that the Brill wave corresponding 
to \eqref{beta_function} contributes to the overall ADM mass to such a degree 
that reduction was not possible, even in the extreme setup of \cite{Dou&Rinn2014} 
where a very wide gluing region starting close to the black hole horizons 
was chosen.

\section*{References}

\bibliography{proceedings} 

\providecommand{\newblock}{}
\begin{thebibliography}{10}
\expandafter\ifx\csname url\endcsname\relax
  \def\url#1{{\tt #1}}\fi
\expandafter\ifx\csname urlprefix\endcsname\relax\def\urlprefix{URL }\fi
\providecommand{\eprint}[2][]{\url{#2}}

\bibitem{Corvino2000}
Corvino J 2000 {\em Commun. Math. Phys.\/} {\bf 214} 137

\bibitem{Corvino2006}
Corvino J and Schoen R~M 2006 {\em J.~Diff.~Geom.\/} {\bf 73} 185
  (\emph{Preprint} \href{http://arxiv.org/abs/gr--qc/0301071}{gr--qc/0301071})

\bibitem{Cortier2013}
Cortier J 2013 {\em Ann. Henri Poincar\'e\/} {\bf 14} 1109 (\emph{Preprint}
  \href{http://arxiv.org/abs/arXiv:1202.3688}{1202.3688 [gr--qc]})

\bibitem{Chrusciel2008}
Chru\'sciel P~T and Pollack D 2008 {\em Ann. Henri Poincar\'e\/} {\bf 9} 639
  (\emph{Preprint} \href{http://arxiv.org/abs/arXiv:0710.3365}{0710.3365
  [gr--qc]})

\bibitem{Chrusciel2009}
Chru\'sciel P~T and Delay E 2009 {\em Comm.~Anal.~Geom.\/} {\bf 17} 343
  (\emph{Preprint} \href{http://arxiv.org/abs/arXiv:0711.1557}{0711.1557
  [gr--qc]})

\bibitem{Giu&Hol2005}
Giulini D and Holzegel G 2005 Corvino's construction using {B}rill waves
  \emph{{P}reprint} \href{http://arxiv.org/abs/gr-qc/0508070}{gr-qc/0508070}

\bibitem{Dou&Rinn2014}
Doulis G and Rinne O 2014 Numerical construction of initial data for
  {E}instein’s equations with static extension to space-like infinity
  \emph{{P}reprint} \href{http://arxiv.org/abs/arXiv:1411.7878}{1411.7878
  [gr-qc]}

\bibitem{Chrusciel2002}
Chru\'sciel P~T and Delay E 2002 {\em Class.\ Quantum Grav.\/} {\bf 19} L71
  (\emph{Preprint} \href{http://arxiv.org/abs/gr--qc/0203053}{gr--qc/0203053})

\bibitem{Rinne2010}
Rinne O 2010 {\em Class.\ Quantum Grav.\/} {\bf 27} 035014 (\emph{Preprint}
  \href{http://arxiv.org/abs/arXiv:0910.0139}{0910.0139 [gr--qc]})

\bibitem{Friedrich1998}
Friedrich H 1998 {\em J.\ Geom.\ Phys.\/} {\bf 24} 83

\bibitem{Beyer2012}
Beyer F, Doulis G, Frauendiener J and Whale B 2012 {\em Class.\ Quantum
  Grav.\/} {\bf 29} 245013 (\emph{Preprint}
  \href{http://arxiv.org/abs/arXiv:1207.5854}{1207.5854 [gr--qc]})

\bibitem{Doulis2013}
Doulis G and Frauendiener J 2013 {\em Gen.\ Relativ.\ Gravit.\/} {\bf 454} 1365
  (\emph{Preprint} \href{http://arxiv.org/abs/arXiv:1301.4286}{1301.4286
  [gr--qc]})

\bibitem{Bril&Linq1963}
Brill D~R and Lindquist R~W 1963 {\em Phys.\ Rev.\/} {\bf 131} 471

\bibitem{Brill1959}
Brill D~R 1959 {\em Ann.~Phys.\/} {\bf 7} 466

\end{thebibliography}

\end{document}